\documentclass[prl,twocolumn,amsmath,amssymb,groupedaddress]{revtex4}

\usepackage{graphicx,subfigure}
\usepackage{bm}
\usepackage{verbatim}
\usepackage{amsmath}
\usepackage{amssymb}
\usepackage[T1]{fontenc}
\usepackage{ae,aecompl}
\usepackage{color}
\usepackage{wasysym}

\newcommand{\pt}{{\partial}}

\newcommand{\rv}{{\bf r}}

\newcommand{\bv}{{\bf b}}

\newcommand{\zh}{{\hat{\bf z}}}

\newcommand{\oh}{{\frac{1}{2}}}

\newcommand{\cH}{{\mathcal H}}
\newcommand{\cL}{{\mathcal L}}
\newcommand{\grad}{{\bm{\nabla}}}

\newcommand{\vsigma}{{\bm{\sigma}}}
\newcommand{\vtau}{{\bm{\tau}}}

\newcommand{\be}{\begin{equation}}
\newcommand{\ee}{\end{equation}}
\newcommand{\bea}{\begin{eqnarray}}
\newcommand{\eea}{\end{eqnarray}}
\newcommand{\bse}{\begin{subequations}}
\newcommand{\ese}{\end{subequations}}
\def\rf#1{(\ref{#1})}

\begin{document}
\title{Fractons from vector gauge theory}
\author{Leo Radzihovsky}
\affiliation{
Department of Physics and Center for Theory of Quantum Matter\\
University of Colorado, Boulder, CO 80309}
\author{Michael Hermele}
\affiliation{
Department of Physics and Center for Theory of Quantum Matter\\
University of Colorado, Boulder, CO 80309}

\date{February 25, 2019}
\email{radzihov@colorado.edu}

\begin{abstract}
  Motivated by the prediction of fractonic topological defects in a
  quantum crystal,
  we utilize a reformulated elasticity duality to derive a description of a
  fracton phase in terms of coupled {\em vector} U(1) gauge theories.
  The fracton order
  and restricted mobility emerge as a result of an unusual Gauss law
  where electric field lines of one gauge field act as sources of charge 
  for others.
  At low energies this vector gauge theory reduces to the
  previously studied fractonic symmetric tensor gauge theory. We
  construct the corresponding lattice model and a number of
  generalizations, which realize fracton phases via a condensation of
  string-like excitations built out of charged particles, analogous to the
  $p$-string condensation mechanism of the gapped 
  X-cube fracton phase.
\end{abstract}

\maketitle

\noindent{\em Introduction.}
Motivated by continued interest in topological quantum matter and by a
search of fault-tolerant quantum memory, recent studies have led to
fascinating developments in an exotic class of quantum spin-liquid
models\cite{chamon,bravyi,Haah,fracton1,fracton2,slagle,HNreview}. These
are characterized by many nontrivial properties, the most unusual of
which are system-size-dependent ground state degeneracy  and the
existence of quasi-particles, dubbed ``fractons'', that exhibit
restricted mobility. Namely, there are quasi-particles confined to
zero-, one- and/or two-dimensional subspaces of the full
three-dimensional space of the model.  While such fracton phases were
originally discovered in fully gapped phases of commuting projector
lattice spin Hamiltonians, it was more recently pointed out
\cite{Pretko_sub}, that fractonic charges are also realized in gapless
phases of U(1) symmetric tensor gauge theories \cite{CenkeXu}.

In a parallel development, it was observed by one of us (L.R.)
\cite{conjectureLR} that such restricted quasi-particle mobility is
strongly reminiscent of the immobile disclinations and glide-only
dislocations in an ordinary two-dimensional (2D) crystal, described by
a symmetric strain tensor field. This conjecture of fracton-elasticity duality 
was reported and moreover explicitly demonstrated in \cite{PretkoLRdualityPRL2018},
utilizing a generalization of boson-vortex duality \cite{DasguptaHalperin,FisherLee}. It was shown that
a 2+1D quantum crystal is dual to a symmetric tensor gauge theory,
with disclinations and dislocations mapping onto fractonic charges and
their dipolar bound states, and with stress tensor $\sigma_{ik}$ and
momentum vector $\pi_k$ fields respectively corresponding to the
electric tensor $E_{ik}$ and magnetic vector $B_k$ fields.

\noindent{\em Motivation and results.}  An important source of insight
into fracton physics has been to relate apparently exotic fracton
states to more familiar quantum phases of matter.  Indeed, the
fracton-elasticity duality is an example of such a relationship.
Related progress has also been made for certain gapped fracton phases,
via a construction of these phases in terms of coupled layers of
ordinary 2D topologically ordered states \cite{ma17coupled,
  vijay17coupled}.  So far there is a relative paucity of
relationships between gapless fracton phases and better understood
phases or theories.  Most U(1) tensor gauge theories are not dual to
elasticity, and even for those that are, developing alternative
viewpoints is highly desirable.
 
Remarkably, the fracton-elasticity duality itself contains the seed of
another such point of view.  There is a sense in which elasticity,
formulated in terms of a symmetrized strain tensor $u_{ik} =
\oh(\partial_i u_k + \partial_k u_i)$ is a system of two ``spin''
flavors of XY models, joined together via ``spin-space'' coupling, as
in systems with spin-orbit interaction.  However, in the absence of
spin-space coupling, such a system dualizes to two independent flavors
of U(1) vector gauge theory, and lacks fractonic charges.  It is thus
natural to ask whether fractonic tensor gauge theories can be
formulated in terms of coupled vector gauge theories and if so, what
minimal ingredients are required for such coupling.  Some progress has
been made along these lines from a different point of view, starting
with a vector U(1) gauge theory and gauging certain global symmetries
to obtain a fractonic tensor gauge theory
\cite{WilliamsonVectorFractons}.  We make contact with this result
below in greater detail.

In this Letter, we first utilize a reformulated fracton-elasticity
duality to derive a 2+1D U(1) {\em vector} gauge theory that hosts
fractonic charges and is equivalent at low energy to the rank-2
symmetric tensor U(1) gauge theory with scalar charge, dubbed the
``scalar-charge theory''.  We then discuss a lattice version of the
same theory, which allows the scalar-charge theory to be understood
starting from two decoupled vector U(1) gauge theories, and condensing
certain charged loops.  Finally, we discuss generalizations of our
lattice construction that provide constructions of a new class of
fractonic tensor gauge theories as coupled vector gauge theories.

The continuum Hamiltonian density we obtain is given by
\begin{eqnarray}
\label{eqn:mainL}
  \tilde{\cH} &=&
  \oh C |{\bf  E}_k|^2 + \oh(\grad\times {\bf  A}_{k})^2
  +\oh K |{\bf  e}|^2\nonumber\\
  &&+ \oh (\grad\times{\bf  a} +  A_a)^2 
  -{\bf  A}_{k}\cdot{\bf J}_{k} - {\bf a}\cdot{\bf j}\; .\;\;\;
\end{eqnarray}
which involves three U(1) vector gauge fields with electric fields
${\bf E}_k$ (with flavors $k = x,y$) and ${\bf e}$ and corresponding
canonically conjugate vector potentials ${\bf A}_k$ and ${\bf a}$.  We
denote the corresponding charge densities by $p_k$ and $\rho$, and
currents by ${\bf J}_k$ and ${\bf j}$, with $p_k$ and ${\bf J}_k$
referred to as the dipole charge and dipole current, respectively.
Moreover, $ A_a = \epsilon_{ik} A_{ik}$, and we use a short-hand
notation where the curl of a 2D vector field is implicitly its scalar
z-component, \emph{i.e.}, $\grad\times{\bf a}\equiv\zh\cdot\grad\times{\bf
  a}$.    
  
The Hamiltonian is supplemented by the Gauss' law constraints:
\begin{eqnarray}
\grad\cdot{\bf E}_k &=& p_k -  e_k,\label{Coulomb_p}\\
\grad\cdot{\bf  e}&=& \rho \text{.} \label{Coulomb_rho}
\end{eqnarray}
Crucially, the components of the electric field $ e_k$ appear as
additional dipole charge in the Gauss's law \rf{Coulomb_p}.

The generalization to $d$ dimensions is straightforward and consists
of $d+1$ U(1) gauge fields obeying the same Gauss' laws but with $k =
1,\dots, d$.  The main difference in the Hamiltonian is that the
$(\grad\times{\bf a} + A_a)^2$ term is replaced by a sum of the
$d(d-1)/2$ terms of the form $(\partial_i a_j - \partial_j a_i + A_{i
  j} - A_{j i})^2$.  The resulting theory is equivalent at low energy
to the $d$-dimensional scalar-charge theory, as we will detail below.


The fractonic nature of the $\rho$ charges can be seen by observing
that moving such a charge requires creating or destroying field lines
of the ${\bf e}$ electric field, but since these field lines
themselves carry gauge charge, a single ``piece'' of field line cannot
be locally created or destroyed.  The immobility of these charges is
also manifest in that gauge invariance requires the current ${\bf j}$
to vanish identically, as we elaborate below. We next turn to the
derivation of this fractonic coupled {\em vector} gauge theory and its
connection to the previously studied tensor scalar-charge theory.


\noindent{\em Derivation.} 
To this end, we pass to Lagrangian formalism and 
begin with an elastic theory of a 2+1D quantum crystal formulated
in terms of  
the phonon field
$u_k$ and its canonically conjugate momentum $\pi_k$. For
simplicity we take the elastic tensor $C_{ij,kl}$ to be
$C_{ij,kl}=C\delta_{ik}\delta_{jl}$. The generalization to an
arbitrary $C_{ij,kl}$ is straightforward.


Ref.\onlinecite{PretkoLRdualityPRL2018} started with such a theory
written in terms of the symmetrized strain $u_{i k} = \partial_i u_k
+ \partial_k u_i$, and showed it is dual to a symmetric tensor gauge
theory, where fractonic charges correspond to disclinations and
dipoles to dislocations.  To get to an equivalent flavored {\em
  vector} gauge theory description, we reformulate the elastic theory
in terms of ``minimally''-coupled quantum XY models, introducing the
orientational bond-angle field, $\theta$ and its canonically conjugate
angular momentum density $L$. The Lagrangian density is given by
\begin{eqnarray}
\cL &=&  \pi_k\partial_t u_k + L\partial_t\theta - \oh\pi_k^2 - \oh C(\pt_i  u_k -   \theta\epsilon_{ik})^2
\nonumber \\ 
&-& \oh  L^2 - \oh K(\grad\theta)^2. \label{Lutheta}
\end{eqnarray}
Due to the coupling to $\theta$, the anti-symmetric part of the unsymmetrized strain $\pt_i  u_k$
is massive, below a scale set by $C$, similar to the Higgs mechanism for gauge fields. Integrating out
$\theta$ results in the standard elasticity theory formulated in terms of the symmetrized strain $u_{i k}$, which is the starting point  of Ref.\onlinecite{PretkoLRdualityPRL2018}.
  To proceed, it is convenient to decouple
the elastic and orientational terms in (\ref{Lutheta}) via Hubbard-Stratonovich vector
fields, stress $\vsigma_{k}$ and torque $\vtau$, resulting in the Lagrangian density
\begin{eqnarray}
\label{LuthetaQ}
\cL
&=& \pi_k\partial_t u_k + L\partial_t\theta 
- \vsigma_{k}\cdot\grad u_k + \sigma_a\theta - \vtau\cdot\grad\theta
\nonumber\\
&& + \oh C^{-1}\vsigma_{k}^2 - \oh\pi_k^2 
- \oh L^2 + \oh K^{-1}\vtau^2,
\end{eqnarray}
where $\sigma_a\equiv \epsilon_{ik}\sigma_{ik}$.



For a complete description, in addition to the single-valued (smooth)
Goldstone mode degrees of freedom, $\theta^e$ and $u^e_k$, we must
also include topological defects.  A disclination defect is defined by
a nonsingle-valued bond angle with winding $\oint d\theta^s = 2\pi
s/n$ around the disclination position, or equivalently in a
differential form, $\grad\times\grad\theta^s = \frac{2\pi
  s}{n}\delta^2(\rv)\equiv\rho(\rv)$.
The integer disclination charge $s$ corresponds to an integer-multiple
of $2\pi/n$ missing (added) wedge of atoms for $s>0$ ($s<0$) in a
$C_n$ symmetric crystal, with most common case of a hexagonal lattice,
$n=6$. 

A dislocation is a point vector defect, around which the displacement
$u_k$ is not single-valued, with winding $\oint d u_k = b_k$, or
equivalently in a differential form, $\grad\times\grad u^s_k =
b_k\delta^2(\rv)\equiv b_k(\rv)$.
An elementary dislocation is a dipole of $\pm 2\pi/n$ disclinations
and is characterized by a 2D Burgers vector charges, $b_k$, that takes
values in the lattice. An edge dislocation corresponds to a ray of
missing or extra lattice sites, with a Burgers vector lying in the 2d
plane of the crystal. A nontrivial configuration of dislocations,
$\bv(\rv)$ can also contribute to a disclination density, given by
$s_b(\rv) = \zh\cdot\grad\times\bv(\rv)$, with a single disclination
corresponding to an end point of a ray of dislocations.

Expressing the phonon and bond-angle fields in terms of corresponding
singular (s) and elastic (e) parts, $u_k = u_k^e + u_k^s$, $\theta =
\theta^e + \theta^s$,
and integrating over the elastic parts, gives the conservation of linear and angular
momentum, $\partial_t\pi_k - \grad\cdot\vsigma_{k} = 0$ and 
$\partial_t L - \grad\cdot\vtau =\sigma_a$.

Expressing linear momentum conservation constraint in terms of dual
magnetic and electric fields, $\pi_k = \epsilon_{kj}B_j$, $\sigma_{ik}
= -\epsilon_{ij}\epsilon_{k\ell}E_{j\ell}$,
leads to the $k$-flavored Faraday equations, $\partial_t B_k +
\grad\times {\bf E}_{k} = 0$.
As in standard electrodynamics, the Faraday law is solved by
$k$-flavored vector ${\bf A}_{k}$ and scalar $A_{0k}$ gauge
potentials, $B_k = \grad\times{\bf A}_{k}$, ${\bf E}_{k} = -\partial_t
{\bf A}_{k} - \grad A_{0k}$.
We emphasize that, in contrast to the {\em symmetric tensor}
approach\cite{PretkoLRdualityPRL2018,PretkoLRdualityEnrichPRL2018},
here, the $k=(x,y)$-flavored {\em vector} gauge field ${\bf A}_{k}$
has components $A_{ik}$ that form an unsymmetrized tensor field.

Using these definitions reduces the conservation of angular momentum to
$\partial_t (L - A_a) - \grad\cdot(\vtau - \zh\times {\bf A}_0) = 0$,
which is then solved by introducing another set of
vector ${\bf a}$ and scalar $a_0$ gauge fields, giving
\begin{eqnarray}
L &=& \grad\times{\bf a} + A_a,\ \
\tau_k = \epsilon_{kj}(\pt_t a_j + \pt_j a_0 - A_{0j}).
\label{solveL}
\end{eqnarray} 
Using these gauge fields to eliminate $\vsigma_{k}$, $\pi_k$, $\vtau$
and $L$, gives an effective Lagrangian density 
\begin{eqnarray}
\label{Leff}
\tilde\cL &=& \oh C^{-1}(\partial_t {\bf A}_k + \grad A_{0k})^2
- \oh(\grad\times {\bf A}_{k})^2\\
&+& \oh K^{-1} (\pt_t a_k + \partial_k a_0 - A_{0k})^2
- \oh(\grad\times {\bf a} + A_a)^2\nonumber \\
&+& {\bf A}_{k}\cdot{\bf J}_{k} -A_{0k}p_k
+{\bf a}\cdot{\bf j} - a_0\rho \text{.} \nonumber
\end{eqnarray}
Here, the dipole charge $p_k$ is given by the dislocation density
$p_k = \epsilon_{l k} b_l=(\zh\times{\bf b})_k$, the fracton charge $\rho$ is the
disclination density, and the corresponding currents are
given by ${\bf J}_{k} = \epsilon_{lk}\zh\times (\partial_t\grad u_l - \grad\partial_t u_l)$ and 
${\bf j} =  \zh\times(\partial_t\grad\theta - \grad\partial_t\theta)$.
Finally, using Hubbard-Stratonovich transformations
to introduce electric fields canonically conjugate to each vector potential, we obtain (\ref{eqn:mainL}) in Lagrangian form.





The unusual Gauss's law \rf{Coulomb_p} couples three vector gauge
theories, with $k$-th component of the $\bf e$ field acting as an
additional source of charge in the Gauss's law for ${\bf E}_k$. We
also note that taking the divergence on the second index $k$ of the
Gauss's law for ${\bf E}_k$, \rf{Coulomb_p} and using the second law
for ${\bf e}$ to eliminate $\grad\cdot{\bf e}$ from the resulting
right hand side gives,
\begin{eqnarray}
\partial_i\partial_k E_{ik} &=&\tilde\rho.  \label{eqn:gauss}
\end{eqnarray}
This thereby recovers the generalized Gauss's law of scalar-charge
tensor gauge theory, with $\tilde\rho \equiv -\rho + \grad\cdot{\bf
  p}$ the total charge contribution,\cite{PretkoLRdualityPRL2018} that
encodes the additional dipole conservation responsible for immobility
of fractonic charges.\cite{Pretko_sub} We note that, in contrast to
the scalar-charge theory, $E_{i k}$ is not a symmetric tensor, but
effectively becomes symmetric at low energy as we demonstrate below.

We note that the Lagrangian (\ref{Leff}) is invariant under a deformed gauge
transformation,
\begin{eqnarray}
{\bf A}_k &\rightarrow& {\bf A}_k + \grad\chi_k,\ \ \
A_{0k} \rightarrow A_{0k} - \partial_t\chi_k,\\
a_k &\rightarrow& a_k + \pt_k\phi - \chi_k,\ \ \
a_0 \rightarrow a_0 - \pt_t\phi,
\label{unusualGauge}
\end{eqnarray}
with \rf{unusualGauge} ensuring that $\grad\times {\bf a} + A_a$ is
gauge invariant.
Under the $\chi_k$ gauge transformation, the current
source terms in \rf{Leff} shift by $- \chi_k j_k + \pt_t\chi_k\tilde p_k + \grad\chi_{k}\cdot{\bf J}_{k}$,
where $\tilde p_k = p_k - e_k$ is the effective dipole density, that
is a combination of microscopic dipoles and electric
field generated by pairs of fracton charges.
Requiring gauge invariance then leads to the dipole continuity
equation $\pt_t\tilde p_k + \grad\cdot{\bf J}_{k} = -j_k$,
where dipole conservation is violated by a nonzero fracton current ${\bf j}$.  
It follows that in the absence of gapped dipoles, ${\bf j} = 0$ for on-shell processes,
\emph{i.e.} isolated disclinations are immobile fractonic charges.

The harmonic \endnote{In a fully nonlinear elasticity, the
    Hamiltonian is an expansion in the {\em nonlinear} strain tensor,
    $u_{ij} = \oh(\pt_i{\bf R}\cdot\pt_j{\bf R}-\delta_{ij})=\oh(\pt_i
    u_j + \pt_ju_i + \pt_i{\bf u}\cdot\pt_j{\bf u})$, that, by
    construction is a target-space ${\bf R}(r_i)$ scalar, i.e.,
    invariant under rotation of ${\bf R}\rightarrow{\bf R'} =
    O\cdot{\bf R}$. This corresponds to $u'_k =
    (O_{kj}-\delta_{kj})x_j + O_{kj} u_j$, that, in a linear
    approximation (small rotation angle $\beta$) reduces to
    (\ref{transrot}).} elasticity theory (\ref{Lutheta}) enjoys the
symmetries
\begin{eqnarray}
u_k &\to& u_k + \alpha_k + \beta \epsilon_{k j} r_j,\label{transrot}\\
\theta &\to& \theta - \beta \text{.} \nonumber
\end{eqnarray}
The constant shift of $u_k$ by $\alpha_k$ can be interpreted as
continuous translational symmetry.  The terms proportional to $\beta$
are a small-angle rotation, where the displacements $u_k$ in the
initial configuration (before symmetry transformation) are also small.
By introducing background gauge fields for these symmetries (see
Appendix) and carrying out the duality in the presence
of the background fields, we identify corresponding conserved currents
on the gauge theory side.  Associated with the $\alpha_k$
translational symmetry, we have the linear momentum current $J^m_{\mu
  k}$, where the conserved density $J^m_{0 k} = \epsilon_{k j} B_j$ is
the magnetic flux, and $J^m_{i k} = \epsilon_{k j} \epsilon_{i \ell}
E_{\ell j}$.
Operators transforming under $\alpha_k$ are thus monopole operators of the ${\bf A}_k$ gauge fields.  Associated with the $\beta$ rotation symmetry and conservation of angular momentum, we have the magnetic flux current of the ${\bf a}$ gauge field, $j^m_{\mu}$.  Due to the coupling between the vector gauge fields, the naive magnetic flux  $\nabla \times {\bf a}$ is not gauge-invariant, requiring modified expressions $j^m_0 = \epsilon_{i j} (\partial_i a_j + A_{i j}) - r_i \epsilon_{i j} J^m_{0 j}$ and $j^m_j = \epsilon_{i j} (\partial_i a_0 + \partial_t a_i - A_{0 i} ) - r_i \epsilon_{i k} J^m_{j k}$.  We note that the current $j^m_{\mu}$ is explicitly position-dependent, similar to the symmetries discussed in \cite{seiberg19vector}. If the position-dependent terms are dropped, $j^m_{\mu}$ remains gauge-invariant but is no longer conserved.  This identification of magnetic flux currents in the coupled vector gauge theory with symmetries of the elasticity theory is useful in our discussion of lattice models below.

We conclude by demonstrating that in fact this dual coupled {\em
  vector} U(1) gauge theory, at low energies is indeed equivalent to
the {\em symmetric tensor} gauge theory. To this end, we observe that
the enlarged gauge redundancy allows us to completely eliminate $a_k$
from the Lagrangian (\ref{Leff}), by choosing $\chi_k = a_k$. The term $\oh(\grad\times {\bf a} + A_a)^2$
reduces to $\oh A_a^2$, thereby gapping out the antisymmetric
component $A_a = \epsilon_{ij}A_{ik}$. Thus, at energies well below
this gap, $\epsilon_{ik}A_{ik}\approx0$, and only the symmetric components of
$A_{ik}$ remain as active degrees of freedom. Furthermore, the electric field term
reduces to $\oh K^{-1}(\pt_t{\bf a} + \grad a_0 - A_{0k})^2\rightarrow
\oh K^{-1}(\grad a_0 - A_{0k})^2$, enforcing $A_{0k}=
\partial_k\alpha_0$. Thus at low energies, this reduces the Lagrangian
exactly to that of the symmetric tensor gauge theory, with the Gauss' law (\ref{eqn:gauss}). 

\emph{Lattice model and charged loop condensation}.  We now consider a 
lattice version of the Hamiltonian (\ref{eqn:mainL}).
  For simplicity of presentation we 
first work in 2D, then discuss the generalization to arbitrary dimension.  The 3D version of the model
appeared previously in \cite{WilliamsonVectorFractons}. We use the resulting model to obtain
a physical picture of the scalar-charge theory in terms of
condensation of certain charged loops. This differs in perspective from the results of \cite{WilliamsonVectorFractons}, where the fracton phase arises upon gauging certain global symmetries.
There are subtleties particular to the 2D case that we discuss, having to do with
the role of compact gauge fields and associated symmetries.

\begin{figure}[t]
	\includegraphics[width=0.8\columnwidth]{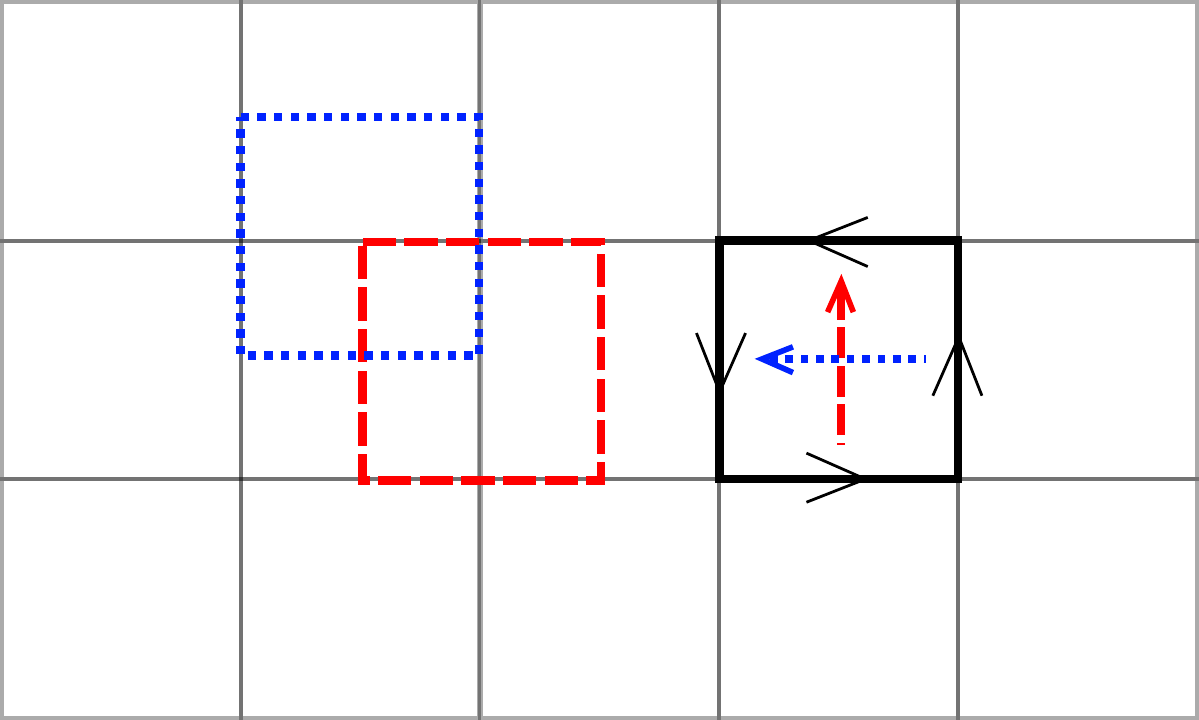}
	\caption{Geometry of the lattice coupled vector gauge theory
          Hamiltonian (\ref{eqn:latticeH}).  Gray solid lines
          represent the underlying square lattice.  On the left, a
          single plaquette is shown for the $x$-lattice (dashed lines;
          red online) and the $y$-lattice (dotted lines; blue online).
          On the right, the thick solid line represents a loop of
          ${\bf e}$ electric field on a plaquette of the underlying
          lattice, with the arrows indicating the direction of ${\bf
            e}$.  Because the $e_k$ electric field lines carry gauge
          charge of the ${\bf E}_k$ electric field as expressed by the
          Gauss law (\ref{Coulomb_p}), there are also necessarily
          lines of ${\bf E}_x$ (dotted line with arrow; blue online)
          and ${\bf E}_y$ (dashed line with arrow; red online)
          electric field.}
	\label{fig:redbluelattice}
\end{figure}

  The lattice geometry consists
of three interpenetrating 2D square lattices as shown in
Fig.~\ref{fig:redbluelattice}.  One of these we refer to as the
``underlying lattice,'' and the $k$-lattice ($k = x,y$) is a square
lattice with vertices the $k$-directed links of the underlying
lattice.  We place the electric field ${\bf  e}$ and vector
potential ${\bf  a}$ on the links of the underlying lattice, with
${\bf E}_k, {\bf A}_k$ placed on the links of the $k$-lattice.  The
${\bf  e}, {\bf  a}$ gauge field is taken to be compact, thereby allowing 
a loop-condensation phase transition into the fracton phase, with
${\bf  e}$ taking integer eigenvalues and ${\bf  a}$ a
$2\pi$-periodic phase.  For simplicity (apart from 2D),
 ${\bf E}_k$ and ${\bf A}_k$ are taken
non-compact, with real eigenvalues.  The Gauss's laws are
$\nabla \cdot \boldsymbol{E}_k = e_k$ and $\nabla \cdot \boldsymbol{e} = 0$,
where the derivatives denote lattice finite differences.  For
simplicity, we do not include any additional charged matter; including it does
not affect the following discussion.

We consider the following lattice Hamiltonian counterpart of (\ref{eqn:mainL}):
\begin{eqnarray}
H &=& \frac{U_{E}}{2}  \sum_{\ell \in L_k} E_{k \ell}^2 + \frac{U_e}{2} \sum_{\ell \in L} e_{\ell}^2 
+ \frac{K_E}{2}   \sum_{\square_k} ( \nabla \times {\bf A}_k)^2 \nonumber \\
&-& K_e \sum_{\square} \cos\big[ ( \nabla \times {\bf a})_{\square} + A_{xy}(\ell_x) - A_{yx}(\ell_y) \big]   \text{.} \label{eqn:latticeH} 
\end{eqnarray}
Here, $k=x,y$ is summed over in those terms where it appears, and $L$
and $L_k$ are the sets of links in the underlying lattice and
$k$-lattice, respectively.  Similarly $\square$ and $\square_k$ denote
plaquettes of the underlying lattice and the $k$-lattice.  The
expression $(\nabla \times {\bf a})_{\square}$ is the lattice line
integral of ${\bf a}$ taken counterclockwise around the perimeter of
the plaquette $\square$. To understand the last term, note that a
single $x$-lattice link $\ell_x$ and single $y$-lattice link $\ell_y$
pass through the center of each plaquette $\square$, as shown on the
left of Fig.~\ref{fig:redbluelattice}.  
This term is thus a gauge-invariant operator that creates electric
field configurations like that shown on the right of
Fig.~\ref{fig:redbluelattice}.

For the generalization to $d$ dimensions, we start with an underlying
hypercubic lattice, and $d$ hypercubic $k$-lattices, with $k=
1,\dots,d$, whose vertices are centered on $k$-directed links of the
underlying lattice.  The $K_e$ operator is replaced with a term
proportional to ${\cos \big[ ( \nabla \times {\bf a})_{\square} + A_{i
    j} - A_{j i} \big]}$, where $\square$ is one of the ${d(d-1)/2}$
types of plaquettes in the underlying lattice, bisected by the two
perpendicular links $\ell_i$ and $\ell_j$.

Considering the 3D case, Ref.~\cite{WilliamsonVectorFractons} showed that one
obtains the Hilbert space of the scalar-charge theory upon taking the
limit $K_e \to \infty$.  We take a different point of view,
considering the phases of the Hamiltonian (\ref{eqn:latticeH}).
When $U_e \gg K_e$, we put the ${\bf  e}$ gauge field into its
confining phase, with ${\bf  e} \approx 0$, and electric field
loops costing an energy proportional to their length.  In this limit
${\bf e}$ can be integrated out and we wind up with $d$ decoupled
${\bf E}_k, {\bf A}_k$ non-compact U(1) vector gauge theories.

Starting from this phase, we increase $K_e$, thus increasing the
fluctuations of the ${\bf e}$ loops.  After $K_e$ is raised above a
critical value, the ${\bf e}$ loops proliferate and condense. This is
analogous to the $p$-string condensation in the coupled layer
construction of the X-cube fracton model \cite{ma17coupled,
  vijay17coupled}, because each ${\bf e}$ loop is built from
point-like charged particles of the ${\bf E}_k$ gauge fields.  We can
access the condensed phase by taking $K_e$ large and expanding the
cosine in the last term. The resulting Gaussian model is simply a
lattice regularization of the continuum Hamiltonian (\ref{eqn:mainL}),
and is identical at low energy to the scalar-charge tensor gauge
theory.

In $d=2$ this picture breaks down, and -- as for a single compact
${\rm U}(1)$ vector gauge theory \cite{polyakov77} -- there is only
one phase.  This occurs because when ${\bf e}$ is compact, we only
have the two U(1) symmetries associated with conservation of $\nabla
\times {\bf A}_k$ magnetic flux, with currents $J^m_{\mu k}$ discussed
above.  When $U_e \gg K_e$, we have two decoupled non-compact U(1)
gauge theories in their deconfined phase, which is of course dual to
the ordered phase of two decoupled XY models, where the two XY
currents are $J^m_{\mu k}$.  When $U_e \ll K_e$, naively one may
expect to be able to expand the cosine to obtain the Hamiltonian
(\ref{eqn:mainL}).  But, as we can see from the dual elasticity
description, the resulting theory is only distinct from two decoupled
XY models in the presence of the \emph{third} $j^m_{\mu}$ U(1)
symmetry, which is absent for compact ${\bf a}$ for any finite
$K_e/U_e$. The absence of a phase transition is obvious on the
elasticity side, corresponding to an incommensurate substrate that
breaks rotational symmetry and thereby gaps out $\theta$ in
(\ref{Lutheta}), reducing to two XY models for $u_x$ and
$u_y$. Therefore there is only one phase in the $d=2$
case. Alternatively we could have taken ${\bf e}$ to be non-compact,
in which case we do have all three U(1) symmetries, but there is still
only one phase.

These considerations have an interesting consequence for the $d=2$
scalar-charge theory.  Namely, if only the two $J^m_{\mu k}$
symmetries are present, there is no difference between this theory and
two non-compact vector U(1) gauge theories.  We note that these
conclusions can be obtained working entirely on the lattice; that is,
we can follow standard techniques to obtain a dual lattice theory of
(\ref{eqn:latticeH}), which at the Gaussian level and in the continuum
limit is identical to elasticity.

\emph{Generalization to vector-charge theory}.  We briefly describe a
generalization of the above construction that reduces to a different
fractonic symmetric-tensor gauge theory at low energy, the {\em
  vector}-charge theory \cite{Pretko_sub}.  In the vector-charge
theory, as with the scalar-charge theory, the electric field
$\varepsilon_{i j}$ and gauge potential $\alpha_{i j}$ are symmetric
tensors, but the Gauss' law constraint is different, given by
$\partial_i \varepsilon_{i j} = \rho_j$, with the gauge charge $\rho_j$
now carrying a vector index.
 
Focusing on 3D and working in the continuum for simplicity, we
introduce a theory of coupled vector gauge theories that reduces to
the vector-charge tensor gauge theory at low energy.  We introduce
\emph{six} ${\rm U}(1)$ gauge fields, with electric fields ${\bf e}_k$
and ${\bf E}_k$ ($k = x,y,z$), and corresponding vector potentials
${\bf a}_k$ and ${\bf A}_k$.  We take the Gauss' law constraints to be
 \begin{eqnarray}
 \partial_i e_{i k} &=& 0 \\
 \partial_i E_{i k} &=& \epsilon_{k i j} e_{i j} \text{,} \label{eqn:vect-gauss}
 \end{eqnarray}
 which express that the anti-symmetric components of $e_{i j}$ act as sources
 of gauge charge for the ${\bf E}_k$ electric fields.
 These constraints are encoded in the Lagrangian density
 \begin{eqnarray}
{\cal L}  &=& - e_{i k} (\partial_t a_{i k} + \partial_i a_{0 k} 
+ \epsilon_{l i k} A_{0 l} )
- E_{i k} ( \partial_t A_{i k} + \partial_i A_{0 k} ) \nonumber  \\
&-&  \frac{C}{2} {\bf E}_k^2 - \frac{1}{2} (\nabla \times {\bf A}_k)^2 - \frac{K}{2} {\bf e}_k^2 - \frac{1}{2} \sum_{i j} b_{i j}^2 \text{,} \label{eqn:vect}
\end{eqnarray}
where $b_{i j} = \epsilon_{i k l} \partial_k a_{l j} + A_{j i} -
\delta_{i j} A_{k k}$.  In the Appendix, we show that
this theory reduces at low energy to the vector-charge theory, and
discuss how to carry out the construction on the lattice.  We also briefly
remark on a generalization to the 2D vector charge theory. The
Appendix  also discusses some further generalizations of
the lattice coupled vector gauge theory construction of the
scalar-charge theory, that reduce at low energy to the $(m,n)$
theories of \cite{bulmash18Higgs}, and the version of the
scalar-charge theory where the electric field is a traceless,
symmetric tensor \cite{Pretko_sub}.

In summary, motivated by the fracton-elasticity
duality\cite{conjectureLR,PretkoLRdualityPRL2018,PretkoLRdualityEnrichPRL2018},
we utilized its reformulation to derive a fractonic coupled U(1) {\em
  vector} gauge theory representation in terms of $d+1$-coupled gauge
fields, where components of one type of electric field act as charges
for the remaining $d$ gauge theories. At low energies this vector
description is identical to fractonic scalar-charge tensor gauge
theory. We used a lattice version of this model to discuss fracton
order in terms of proliferation of electric field loops. We also
proposed a number of generalizations of this construction, making
contact with fractonic tensor gauge theories that are not dual to
elasticity.

\emph{Acknowledgments}.  We thank Michael Pretko for discussions. MH also acknowledges
discussions with Nathan Seiberg. The
work of LR was supported by the Simons Investigator Award from the
Simons Foundation, and by the Soft Materials Research Center under NSF
MRSEC Grants DMR-1420736.  The work of MH is supported by the
U.S. Department of Energy, Office of Science, Basic Energy Sciences
(BES) under Award number DE-SC0014415, and also partly supported
by the Simons Collaboration on Ultra-Quantum Matter,
which is a grant from the Simons Foundation (651440).

\appendix

\section{Appendix}
\label{appendix}

\emph{Background gauge fields for translation and rotation symmetries}.  To identify the  $J^m_{\mu k}$ linear momentum and the $j^m_{\mu}$ angular momentum currents on the gauge theory side, we introduce background gauge fields ${\cal A}_{\mu k}$ and ${\cal B}_{\mu}$ for the $\alpha_k$ and $\beta$ symmetries of Eq.~(11) in the main text, respectively.  The elasticity Lagrangian with background fields included is
\begin{eqnarray}
\cL &=&  \pi_k( \partial_t u_k + {\cal A}_{0 k} + {\cal B}_0 \epsilon_{k j} r_j )  + L ( \partial_t\theta - {\cal B}_0 ) - \frac{1}{2} \pi_k^2  \nonumber \\
&-&  \frac{1}{2} C(\partial_i  u_k -   \theta\epsilon_{ik} - {\cal A}_{i k} - {\cal B}_i \epsilon_{k j} r_j )^2  \nonumber \\
&-& \frac{1}{2} L^2 - \frac{1}{2} K(\partial_i \theta + {\cal B}_i )^2.
\end{eqnarray}
This is invariant under Eq.~(11) of the main text with $\alpha_k$ and $\beta$ now space and time dependent, combined with the transformations of the background gauge fields,
\begin{eqnarray}
{\cal A}_{0 k} &\to& {\cal A}_{0 k} - \partial_t \alpha_k \nonumber \\
{\cal A}_{j k} &\to& {\cal A}_{j k} + \partial_j \alpha_k \\
{\cal B}_0 &\to& {\cal B}_0 - \partial_t \beta \nonumber \\
{\cal B}_j &\to& {\cal B}_j + \partial_j \beta \text{.} \nonumber
\end{eqnarray}
  From this starting point, it is straightforward to carry through the duality as in the main text to identify the conserved currents on the gauge theory side.  We note that for this purpose, the singular parts of the elastic and bond angle fields can be ignored.

\emph{Coupled vector gauge theory construction of the vector-charge theory.}  Here we show that the coupled vector gauge theory Lagrangian in Eq.~(15) of the main text reduces to the 3D vector-charge theory at low energy, and briefly describe the corresponding lattice model.  We also briefly remark on the generalization of this construction to the 2D vector charge theory.  

The Lagrangian is invariant under the gauge transformations
 \begin{eqnarray}
 a_{i j} &\to& a_{i j} + \partial_i \phi_j + \epsilon_{k i j} \chi_k , \qquad a_{0 j} \to a_{0 j} - \partial_t \phi_j\;\;\;\;\;\; \\
 A_{i j} &\to& A_{i j} + \partial_i \chi_j , \qquad A_{0 j} \to A_{0 j} - \partial_t \chi_j \text{.}
 \end{eqnarray}
 
We would like to eliminate the antisymmetric part of $a_{i j}$ from the Lagrangian.  To do this, we write
\begin{equation}
a_{i j} = \frac{1}{2} a^s_{i j} + \epsilon_{i j k} \Omega_k \text{,}
\end{equation}
where $a^s_{i j} = a_{i j} + a_{j i}$, and using the fact that an antisymmetric rank-2 tensor can be expressed in the form of the second term.  We have
\begin{equation}
\partial_t a_{i j} + \partial_i a_{0 j} + \epsilon_{k i j} A_{0 k}
= \frac{1}{2} \partial_t a^s_{i j} + \partial_i a_{0 j} + \epsilon_{k i j} (A_{0 k} + \partial_t \Omega_k ) \text{,}
\end{equation}
and
\begin{equation}
b_{i j} = \frac{1}{2} \epsilon_{i k l} \partial_k a^s_{l j} + (A_{j i} - \partial_j \Omega_i) - \delta_{i j} (A_{kk} - \partial_k \Omega_k) \text{.}
\end{equation}
Then clearly if we make a partial gauge transformation, where we only transform $A_{i j}$ using $\chi_i = \Omega_i$, and do not transform $a_{i j}$, we eliminate the antisymmetric part of $a_{i j}$.  Integrating out the electric fields ${\bf e}_k$ and ${\bf E}_k$, the resulting Lagrangian is
 \begin{eqnarray}
 {\cal L} &=& \frac{1}{2 K} \sum_{i j} ( \frac{1}{2} \partial_t a^s_{i j} + \partial_i a_{0 j} + \epsilon_{k i j} A_{0 k} )^2
 \nonumber \\ 
&+& \frac{1}{2 C} \sum_{i j} ( \partial_t A_{i j} + \partial_i A_{0 j} )^2 - \frac{1}{2} (\nabla \times {\bf A}_k)^2 \nonumber \\
 &-&  \frac{1}{2} \sum_{i j} ( \frac{1}{2} \epsilon_{i k l} \partial_k a^s_{l j} + A_{j i} - \delta_{i j} A_{k k} )^2 \text{.}
 \end{eqnarray}
 
It is convenient to define
 \begin{equation}
 c_{i j} = \frac{1}{2} \epsilon_{i k l} \partial_k a^s_{l j} \text{,}
 \end{equation}
 and observe $c_{i i} = 0$.  Under a gauge transformation,
 $c_{i j} \to c_{i j} + \frac{1}{2} \epsilon_{i k l} \partial_k \partial_j \phi_l$, and
we define the gauge invariant quantity
 \begin{equation}
 \beta_{i j} = 2 \epsilon_{j k l} \partial_k c_{i l} = \epsilon_{i m n} \epsilon_{j k l} \partial_m \partial_k a^s_{n l} \text{,}
 \end{equation}
 which will play the role of the magnetic field tensor in the low-energy vector-charge theory.
 
We proceed by making the change of variables $A_{i j} \to A_{i j} - c_{j i}$, where the position of the indices should be carefully noted.  We observe that
\begin{equation}
(\nabla \times {\bf A}_k)_i \to (\nabla \times {\bf A}_k)_i - \frac{1}{2} \beta_{k i} \text{.}
\end{equation}
The Lagrangian becomes
 \begin{eqnarray}
 {\cal L} &=& \frac{1}{2 K} \sum_{i j} ( \frac{1}{2} \partial_t a^s_{i j} + \partial_i a_{0 j} + \epsilon_{k i j} A_{0 k} )^2   \\
 &+& \frac{1}{2 C} \sum_{i j} ( \partial_t A_{i j} + \partial_i A_{0 j} - \partial_t c_{j i} )^2 - \frac{1}{2} (\nabla \times {\bf A}_k)^2  \nonumber \\
 &-&   \frac{1}{8} \sum_{i j} \beta_{i j}^2 + \frac{1}{2} \beta_{k i} (\nabla \times {\bf A}_k)_i 
 - \frac{1}{2} \sum_{i j} ( A_{j i} - \delta_{i j} A_{k k} )^2 \text{.} \nonumber
 \end{eqnarray}
 
 We want to integrate out the massive $A_{0k}$ and $A_{i j}$ fields,
 and neglect all their subdominant higher-derivative terms, that are
 unimportant for the long-distance, low-energy behavior. The generated
 leading terms in $a^s_{i j}$ have either two time derivatives, or
 four spatial derivatives, resulting in the dynamical critical
 exponent $z=2$ scaling expected for the vector-charge theory.  All of
 the $c_{j i}$ terms proportional to $1/C$ lead to subdominant
 contributions, as does the second term on the last line, and we drop
 these to obtain
   \begin{eqnarray}
 {\cal L} &=& \frac{1}{2 K} \sum_{i j} ( \frac{1}{2} \partial_t a^s_{i j} + \partial_i a_{0 j} + \epsilon_{k i j} A_{0 k} )^2 
 - \frac{1}{8} \sum_{i j} \beta_{i j}^2 \nonumber \\
  &-& \frac{1}{2} \sum_{i j} ( A_{j i} - \delta_{i j} A_{k k} )^2 \text{.}
 \end{eqnarray}
 Now integrating out the $A_{i j}$, $A_{0 k}$ gauge fields gives
 \begin{equation}
{\cal L} =  \frac{1}{8 K} \sum_{i j} (  \partial_t \alpha_{i j} + \partial_i a_{0 j}  + \partial_j a_{0 i})^2 
 - \frac{1}{8} \sum_{i j} \beta_{i j}^2 \text{,}
 \end{equation}
 where we have identifed $\alpha_{i j} = a^s_{i j}$.  This is
 precisely the Lagrangian for the vector-charge theory.
 
 To carry out this construction on the lattice, we consider an
 underlying cubic lattice, and three cubic $k$-lattices ($k = x,y,z$),
 whose vertices are the $k$-directed links of the underlying lattice.
 We place the ${\bf e}_k$ electric field on the links of the
 $k$-lattice.  To understand where to place the ${\bf E}_k$ electric
 fields, we refer to the Gauss law, Eq.~(14) of the
 main text, which asserts that $\epsilon_{k i j} e_{i j}$ is the
 density of ${\bf E}_k$ gauge charge.  On the lattice, $\epsilon_{k i
   j} e_{i j}$ naturally resides on the plaquettes of the underlying
 lattice normal to the $k$-direction.  For instance, $\epsilon_{z i j}
 e_{i j} = e_{x y} - e_{y x}$, and this variable naturally resides on
 $xy$-plane underlying lattice plaquettes, because a $y$-link of the
 $x$-lattice crosses an $x$-link of the $y$-lattice at the center of
 each such plaquette.  Therefore we introduce $xy$, $yz$ and $xz$
 cubic lattices, whose vertices lie at plaquette centers of the
 corresponding underlying lattice plaquettes, and we place ${\bf E}_x$
 on the links of the $yz$-lattice, ${\bf E}_y$ on the links of the
 $xz$-lattice, and ${\bf E}_z$ on the links of the $xy$-lattice.  This
 results in the usual lattice vector-charge theory where diagonal
 components of $\varepsilon_{i j}$ reside on sites of the underlying
 lattice, with off-diagonal components residing on underlying lattice
 plaquettes.
 
 Finally, we briefly comment on the generalization of this construction to the 2D vector charge theory. In this case we introduce two ${\bf e}_k$ electric fields ($k = x,y$) and one ${\bf E}$ electric field, for a total of three ${\rm U}(1)$ vector gauge fields.  The Gauss' law constraints are
\begin{eqnarray}
\partial_i e_{i k} &=& 0 \\
\partial_i E_i &=& \epsilon_{i j} e_{i j} \text{.}
\end{eqnarray}
The construction then proceeds as in the 3D case, and one obtains the 2D vector charge theory at low energy.  We note that the gauge-invariant magnetic field operators in the 2D construction are $B = \epsilon_{i j} \partial_i A_j$ and
$b_k = \epsilon_{i j} \partial_i a_{j k} + A_k$, where ${\bf a}_k$ and ${\bf A}$ are the vector potentials conjugate to ${\bf e}_k$ and ${\bf E}$, respectively.

 \emph{Further lattice-model generalizations}.  We briefly state some
 generalizations of the lattice construction of the scalar-charge
 theory, a detailed analysis of which will be presented elsewhere.  We
 focus on three dimensions, and introduce an underlying cubic lattice
 and three cubic $k$-lattices ($k=x,y,z$), whose vertices are centered
 on $k$-directed links of the underlying lattice.  We introduce a
 compact U(1) gauge field on each lattice with electric fields ${\bf
   e}$ and ${\bf E}_k$, writing $E_{ik} \equiv ({\bf E}_k)_i$. We
 consider Gauss's laws of the form
\begin{eqnarray}
\nabla \cdot {\bf e} &=& 0 \\
\sum_{i=x,y,z} M_{i k} \Delta_{i}  E_{i k} &=& e_k \text{,}
\end{eqnarray}
where $\Delta_i$ is the lattice derivative operator and $M$ is a $3
\times 3$ matrix with $m$ in the diagonal entries and $n$ in the
off-diagonal entries, with $m$ and $n$ positive, relatively prime
integers.  $m=n=1$ corresponds to an isotropic Gauss's law for the
${\bf E}_k$ electric fields, with $e_k$ carrying ${\bf E}_k$ electric
charge, while $m \neq n$ makes the Gauss's laws anisotropic.  Starting
from a phase where ${\bf e}$ loops are confined and the three ${\bf
  E}_k$ gauge fields are deconfined, condensing ${\bf e}$ loops
produces the deconfined phase of the $(m,n)$ scalar-charge tensor
gauge theories introduced in [20].  These theories
have a symmetric-tensor electric field $E^s_{i j}$ obeying the Gauss's
law $\sum_{i \leq j} M_{ij} \Delta_i \Delta_j E^s_{i j} = 0$.

This construction can be further generalized by adding another vector
gauge field on the underlying lattice, with electric field
$\boldsymbol{{\cal E}}$, and supplementing the above Gauss's laws with
\begin{equation}
\nabla \cdot \boldsymbol{{\cal E}} = \sum_{i} E_{i i} \text{.}
\end{equation}
Constructing a Hamiltonian for these gauge fields and taking the
coefficients of all the Maxwell terms large, so that all cosines can
be expanded, one obtains the traceless version of the $(m,n)$
scalar-charge theory, which obeys the additional constraint $\sum_i
E^s_{i i} = 0$.

\bibliography{ref}

\end{document}